# Warum wir es für eine gute Idee gehalten haben, eine DACH-Spieledatenbank aufzubauen

Eugen Pfister, Aurelia Brandenburg, Adrian Demleitner und Lukas Daniel Klausner

**Zusammenfassung**
Unser Werkstattbericht gibt Einblick in den Entstehungskontext sowie die zugrundeliegenden methodischen Überlegungen hinter der von den Autor\*innen publizierten Spieledatenbank. Diese wurde kollaborativ erarbeitet und führt digitale Spiele, die in Deutschland, Österreich und der Schweiz bis zum Jahr 2000 entwickelt wurden. In diesem Bericht skizzieren wir neben unseren Ausgangsüberlegungen und den verschiedenen Arbeitsschritten bei der Realisierung außerdem auch, auf welcher Datenbasis die Datenbank aufgebaut und geprüft wurde, was die Ziele des Datenmodells sind und mit welchen Schwierigkeiten wir im Prozess der Erstellung konfrontiert waren. Hiernach ordnen wir den aktuellen Stand der Spieledatenbank ein und geben einen Ausblick auf die weiteren Pläne des Projekts.

**Schlüsselwörter**
Game Studies, Spielgeschichte, *local history*, Datenbank, Datenmodellierung, Digitalisierung, Werkstattbericht

## 1. Einleitung und Motivation

Mit dem wachsenden Interesse von Forschung und Journalismus an digitalen Spielen geht auch ein wachsendes Bewusstsein für ihre Geschichte einher. Hier zeigt sich rasch, dass das „neue" Medium mit – je nach gesetztem Anfangspunkt – mindestens fünfzig Jahren Geschichte gar nicht mehr so jung ist. Es zeigte sich aber auch etwas Anderes: Rasch hat sich eine Art US-amerikanisch-japanische Meistererzählung herauskristallisiert, die bis heute häufig unkritisch wiederholt wird. Sie erzählt eine Geschichte von meist männlichen, weißen Genies und Innovatoren, die oft gegen den Zeitgeist ihre Visionen umgesetzt haben: Von den jungen Studenten des MIT, die nächtens einen PDP-10-Rechner zweckentfremdet haben, um darauf das Spiel *Spacewar!* zu entwickeln, bis hin zum Self-made-Millionär und Enfant terrible Nolan Bushnell, der mit der Gründung von Atari den ersten Videospiel-Boom in den USA ausgelöst hat.

Später liest man vom großen Video Game Crash in den USA, von hunderttausenden Spielecartridges, die verschämt in der Wüste vergraben wurden, vom Aufstieg der japanischen Unternehmen Nintendo und Sega und deren „Console Wars". Diese Geschichte, wie wir sie zum Beispiel in Steven Kents *Ultimate History of Video Games* lesen können (Kent 2001), ist nicht grundsätzlich falsch – sie blendet aber quasi alle Länder abseits der USA, Japans und vielleicht noch Großbritanniens aus. Dabei spürte man in Europa beispielsweise nur wenig vom Crash, weil hier Konsolen nie so verbreitet waren wie in den USA. Und trotzdem wurde fast überall gespielt: auf Jahrmärkten, in Spielhallen und zu Hause. In Europa hatten sich etwa dynamische Entwickler*innen-Netzwerke um die hier populären Mikrocomputer gebildet: um den ZX Spectrum und den Amstrad CPC, vor allem aber um den C64, den Commodore Amiga und MS-DOS-kompatible Computer. Um diese Plattformen herum entstanden in Europa, aber auch in Australien und Neuseeland extrem produktive eigenständige Spielekulturen.

Erst in jüngster Zeit hat sich – auch hier zeigt sich ein ähnlicher Trend in Forschung und Journalismus – der Fokus verstärkt auf die regionale und nationale digitale Spielegeschichte verlagert. Graeme Kirkpatrick hat sich der Videospielkultur im Vereinigten Königreich gewidmet (Kirkpatrick 2015), Alexis Blanchet und Guillaume Montagnon forschen den Ursprüngen des „French Touch" nach (Blanchet und Montagnon 2020) und Melanie Swalwell hat sich ausführlich mit den Ursprüngen der sogenannten *Homebrew Culture* in Australien und Neuseeland beschäftigt (Swalwell 2021). Diese ersten Studien zu einer *local history* digitaler Spiele haben eindrücklich gezeigt, dass es sich hierbei nicht um Randnotizen einer primär US-amerikanisch-japanischen Dynamik gehandelt hat. Jaroslav Švelch konnte in seiner Studie zur frühen Geschichte digitaler Spiele in der Tschechoslowakei nachweisen, dass es in den 1980er-Jahren sogar hinter dem Eisernen Vorhang und gegen alle Widrigkeiten zur Entwicklung einer eigenständigen Kultur digitaler Spiele kam (Švelch 2018).

Durch diese Vorarbeiten wurde uns immer deutlicher bewusst, wie dringend notwendig diese Forschung ist. Um ein Beispiel zu nennen: Will man heute etwa den Erfolg der *Anno*-Reihe besser verstehen, hilft kein Blick in die *Ultimate History of Video Games*. In den US-amerikanischen Videospielchroniken findet das Phänomen „Wirtschaftssimulation und Aufbauspiel" selten bis nie Erwähnung. Zum Glück werden wir gerade Zeug*innen vieler gleichzeitig entstehender und miteinander vernetzter Initiativen zur Erforschung der jeweiligen Geschichte digitaler Spiele in ganz Europa sowie auch in Südamerika, Asien, Afrika und Ozeanien. In Deutschland und Österreich sind gerade mehrere Projekte im Entstehen begriffen, und in der Schweiz wird von 2023 bis 2027 ein großes SNF-Sinergia-Projekt zur Erforschung der lokalen Videospielgeschichte finanziert, an dem vier Universitäten und Hochschulen sowie 20 Forscher*innen beteiligt sind, unter anderem auch drei der vier Autor*innen dieses Werkstattberichts.[1]

Alle diese Projekte stehen zu Beginn vor einem ähnlichen Problem: Man weiß grundsätzlich, oft aus eigener Erfahrung, von Spielen aus dem untersuchten geografischen Raum, es gibt aber nur wenige, unvollständige Listen oder Datenbanken dieser Spiele – und manchmal auch gar keine. Solche – und seien es nur rudimentäre – Datensätze sind aber Grundvoraussetzung für die meisten geschichtlichen Forschungsfragen, auch wenn es keine primär quantitativen Studien sind: *Wie viele Spiele wurden insgesamt ungefähr in welchem Zeitraum entwickelt und wie hat sich diese Zahl verändert? Welche Systeme waren wann wo besonders verbreitet? Welche Genres waren wann wo populär? Wie viele Entwickler\*innen waren zu welchen Zeitpunkten normalerweise beteiligt? Wann und wo kam es zu transnationalen Kooperationen?*

---

1 Siehe auch das Blog zum Projekt unter https://chludens.hypotheses.org/.

Egal, ob man sich für frühe Homebrew-Spiele interessiert, für Spiele, die für den Schulunterricht eingesetzt wurden, oder für das Entstehen der ersten kommerziellen Entwicklungsstudios – um ein Phänomen historisch erfassen zu können, ist es notwendig, zumindest einen rudimentären Überblick über die Datenlage zu haben. Das zeigt sich auch daran, wie erschreckend schnell vergessen wird: Auch in unserem Team hat sich herausgestellt, dass wir alle die Anzahl der Spiele, die bis zum Jahr 2000 entwickelt wurden, dramatisch unterschätzt hatten.

Aus den genannten Gründen war es für uns wichtig, eine Datenbank von Spielen für Deutschland, Österreich und die Schweiz zu erstellen. Im Frühjahr 2023 haben wir, nach etwas über einem Jahr Arbeit, eine erste (noch sehr unvollständige) Version unserer Arbeit in Open Access veröffentlicht (Pfister et al. 2023), gerade um es anderen Forscher\*innen, Journalist\*innen oder auch einfach nur Interessierten in Zukunft leichter zu machen. Wie es dazu kam, von den Herausforderungen und Rückschlägen, aber auch von ersten Entdeckungen wollen wir im Folgenden berichten – beginnend mit der Ausgangslage.

## 2. Quellenbasis

Es wäre vermessen, behaupten zu wollen, dass es bisher gar keine Spieledatenbanken für den deutschsprachigen Raum gegeben hätte. Vielmehr ist es so, dass wir ohne die Vorarbeit von tausenden Beitragenden gar nicht in der Lage gewesen wären, unsere Datenbank aufzubauen. Allerdings war die Ausgangslage gerade für unsere Bedürfnisse unzufriedenstellend: Die englischsprachigen Listen auf Wikipedia[2] geben wirklich nur einen allerersten, sehr begrenzten Einblick in die Entwicklungsgeschichte der drei Länder, insbesondere im Hinblick auf die frühe Geschichte digitaler Spiele.

Weitaus ergiebiger sind da schon einschlägige Spiele-Datenbanken. Eine erste Kategorie von diesen sind große, webbasierte Angebote, namentlich *MobyGames* (http://mobygames.com/) und die *Universal Videogame List* (https://www.uvlist.net/). Beides sind umfassende, crowdgesourcte Datenbanken, die für unser Projekt sehr hilfreich beim Nachschlagen von Daten und Fakten zu bereits identifizierten Spielen waren. Was sie jedoch nicht einfach erlauben, ist die Suche nach Ländermerkmalen der beteiligten Personen oder Firmen. MobyGames, die wahrscheinlich vollständigste internationale Datenbank digitaler Spiele, erlaubt es etwa gar nicht, nach den Ländern der Entwickler\*innen zu filtern oder zu durchsuchen, und auch hier hat sich im Zuge unserer Recherchen gezeigt, dass insbesondere frühe im DACH-Raum entwickelte Spiele oft fehlen. Die deutschsprachige *Online-Games-Datenbank* (OGDB, https://ogdb.eu/) wiederum lässt sich zwar nach Herkunftsland filtern, aber auch hier fehlen viele Spiele aus den 1980er- und 1990er-Jahren. Vor allem aber eignet sich die OGDB nur bedingt für wissenschaftliche und journalistische Recherchen, da sie sich zum einen nicht nach Jahren ordnen lässt und zum anderen alle weiterführenden Angaben zu den Entwicklerstudios, Publishern und allen beteiligten Entwickler\*innen fehlen. Die dänische Datenbank *Play:Right* (https://www.playright.dk/) schließlich lässt sich nach Herkunftsland filtern und dann nach Erscheinungsjahr sortieren, ist aber ebenfalls sehr unvollständig.[3]

---

2 Konkret also „Category:Video games developed in Germany" (https://en.wikipedia.org/w/index.php?title=Category:Video_games_developed_in_Germany), „Category:Video games developed in Switzerland" (https://en.wikipedia.org/wiki/Category:Video_games_developed_in_Switzerland) und „List of video games developed in Austria" (https://en.wikipedia.org/wiki/List_of_video_games_developed_in_Austria).

Dieser Kritikpunkt trifft auch auf eine weitere essenzielle Art von Informationsquellen zu, nämlich jene Datenbanken, welche sich auf einzelne Plattformen bzw. Systeme fokussieren. Beispielhaft aufzuführen wären unter anderem *Hall of Light* (https://hol.abime.net/) und *Lemon Amiga* (https://www.lemonamiga.com/), *C64-Wiki* (https://www.c64-wiki.de/) und *Gamebase 64* (http://www.gamebase64.com/) oder *Atari Mania* (http://www.atarimania.com/). Diese Plattformen sind in ihrer Organisation oftmals an Wikis angelehnt. Das heißt, sie funktionieren dank der freiwilligen Teilnahme ungezählter interessierter Personen und Enthusiast\*innen. Diese investieren ihre Freizeit, um die Plattformen mit Wissen zu füttern. Teilweise schlägt sich das in umfassenden (für die Forschung extrem wertvollen) Datensammlungen nieder: Im C64-Wiki etwa finden sich zu vielen kleinen Spielen neben Beschreibungen auch Anleitungen und Lösungen, Karten, Screenshots, Scans der Cover usw. Aus der Sicht der Wissenschaft leiden diese Plattformen aber unter den gleichen Problemen wie auch die zuvor genannten crowdgesourcten Quellen: Die Provenienz der Daten der einzelnen Einträge sind nicht oder nur schwer nachzuvollziehen.

Es ist uns an dieser Stelle wichtig darauf hinzuweisen, dass es sich bei diesen kritischen Anmerkungen nicht um eine Kritik an der Sammlungsarbeit an sich handelt. Schließlich handelt es sich hierbei um Privatinitiativen, die von der unbezahlten Sammel- und Recherchearbeit Einzelner leben. Sie sind eine wertvolle Ressource und waren sehr hilfreich für unsere Arbeit, eignen sich aber nicht als Datenbanken für die Forschung, auch deshalb, weil sie gar keinen Anspruch auf Vollständigkeit stellen können – vor allem aber, weil sie nicht wissenschaftlich kritisch überprüft wurden.

---

3 Bis inklusive 1999 werden etwa nur 12 österreichische, 168 deutsche und 3 Schweizer Spiele aufgelistet.

Unser Ansatz für die Erstellung unserer eigenen Liste beinhaltete deshalb neben der zweifachen Kontrolle auch die Triangulation von Quellen. Wir nehmen an, dass diese Quellen ihr Wissen auch untereinander teilen und so unter Umständen Falschinformationen wiederholen. Daher haben wir in Ansätzen versucht, die Informationen mit Quellen abzugleichen, welche nicht auf diesen spezialisierten Plattformen verzeichnet waren. Eine interessante Differenz in diesen spezialisierten Plattformen lässt sich an den erweiterten Suchmasken ausmachen. Einige Plattformen, wie zum Beispiel Hall of Light und Atari Mania, hatten tatsächlich Optionen für die Suche nach Land oder Sprache. Andere, etwa Gamebase 64, ließen sich stattdessen danach durchsuchen, wer ein Spiel gecrackt hatte.[4]

Zuletzt haben wir außerdem versucht, die Metadaten zu einzelnen Spielen auch darüber hinaus mit entsprechenden Einträgen in größeren Datenbanken abzugleichen, die nicht oder nur schlecht für unsere Zwecke durchsuchbar waren, aber die Überprüfung von Informationen erlauben – Seiten wie etwa die *Interactive Fiction Database* (IFDB, https://ifdb.org/), bei der es sich ebenfalls um ein Wiki-ähnliches Projekt handelt, die aber nicht auf ein System, sondern auf ein Genre fokussiert ist. Diese Seiten katalogisieren so z. B. allgemein Spiele, sind aber mehr auf eine Veröffentlichung und Archivierung von kleinen Projekten und/oder ihren Metadaten ausgelegt und nicht darauf, Spiele unter speziellen Kriterien zu recherchieren. Die komplexeste Zugriffsmöglichkeit stellte hier noch die Sammlung des *Internet Archives* (https://archive.org/details/classicpcgames) dar, das inzwischen auch im großen Stil ältere Software zum Download oder sogar direkt im Browser zugänglich macht und so teilweise durch Filtermöglichkeiten der Suche oder untereinander verlinkte Einträge die Recherche zu einzelnen Spielen erleichtern kann. Außerdem hostet das Internet Archive schön länger eine umfassende Sammlung an Schriftquellen zu digitaler Spielgeschichte, wie etwa Spielemagazine und Werbeanzeigen, mit denen sich ebenfalls z. B. Informationen aus crowdgesourcten Angeboten gegenprüfen lassen.

Während diese Seiten allerdings teils ausführliche Metadaten aufweisen, sind sie alle primär dafür gedacht, das digitale Objekt eines Spiels oder eines Digitalisats zugänglich zu machen, und nicht dafür, eine nach bestimmten Kriterien geordnete Spielliste aufzustellen. Alle Datenbanken und Seiten, welche wir für unsere Recherchen beigezogen hatten, operieren auf der Ebene des Einzelobjektes. Die Metadaten einzelner Spiel-Einträge waren oft genügend bis ausführlich, wir konnten jedoch mit keiner der Seiten eine für die Forschung brauchbare geordnete Spielliste erstellen. Die Organisation von Wissen und das zugehörige Datenmodell erfolgt also primär über einige wenige normierte Felder wie etwa die*den Ersteller*in eines Eintrags in Kombination mit einem Schlagwort- und/oder Kategorie-Sammlungssystem, das allerdings abhängig von der*dem

---

4 Das ist für uns zwar nicht unmittelbar relevant, zeigt jedoch auf, wie eine Suchmaske etwas über das epistemologische Modell einer Plattform aussagen kann.

Bearbeiter*in eines Eintrags variieren kann, wenn verschiedene User*innen zwar inhaltlich dasselbe meinen, es aber unterschiedlich verschlagworten. Bleibt schließlich eine Ausnahme zu erwähnen: der *Swiss Games Garden* (https://swissgames.garden/). Die Sammlung entstand aus einer Kooperation des Schweizer Entwicklers David Stark und der zwei Forscher David Javet und Yannick Rochat, wobei Letzterer auch Kollege in unserem SNF-Sinergia-Projekt Confederatio Ludens ist. Die Datenbank ist gerade erst im Aufbau begriffen und wir haben bereits dazu beigetragen, sie mit unseren Funden weiter auszubauen. Wir möchten dabei betonen, dass es nie unser Ziel war, eine „wissenschaftliche" Konkurrenz zu bestehenden Datenbanken aufzubauen, sondern unsere Datensammlung so offen zu gestalten, dass sie potenziell allen zur Verfügung steht. Im Swiss Games Garden und in Wikidata (siehe unten) haben wir selbst unsere Daten wiederum eingepflegt.

## 3. Zum Aufbau einer Alpha-Version

Ausgehend von unserer Forschungsfrage und der Quellenbasis gab es nun in unserer Wahrnehmung zwei Möglichkeiten: Der wissenschaftlich saubere Weg wäre es gewesen, in Form eines finanzierten Projekts eine neue Infrastruktur aufzubauen, mit mehreren bezahlten Mitarbeiter*innen im Vorfeld den Aufbau der Datenbank zu strukturieren, ein kontrolliertes Vokabular zu erstellen und dann gezielt mit der Suche und Befüllung der Datenbank anzufangen. Erfahrungsgemäß bräuchte ein solches Projekt eine Vorlaufzeit von mindestens einem Jahr für Recherchen im Vorfeld und das Schreiben des Antrags. Dann bestünde je nach Förderinstrument eine 10- bis 20-prozentige Chance, dass das Projekt auch finanziert würde, woraufhin wir drei bis vier Jahre Zeit hätten, es umzusetzen. Vorteil wäre, dass viele schwierige Fragen im Vorfeld geklärt werden könnten und das Team sehr gezielt und aufeinander abgestimmt bei der Suche vorgehen könnte. Nachteil wäre die lange Vorbereitungszeit des Projekts und dass die Daten dann erst nach vier Jahren zur Verfügung stünden, vor allem aber, dass die Aussichten auf Finanzierung insgesamt gering wären. Was also tun, wenn das Projekt nicht finanziert würde? Wir könnten natürlich das Projekt umschreiben, nachbessern und auf eine Finanzierung im nächsten Jahr hoffen – im besten Fall.

Aus diesen Gründen hat sich Eugen Pfister im Frühjahr 2022 für einen anderen Zugang entschieden. Zu diesem Zeitpunkt hatte er schon Listen an Schweizer und österreichischen Spielen angelegt, ihm fehlten aber wichtige Informationen zu einem der größten Spielmärkte Europas, mit dem die Schweiz und Österreich eng verwoben waren: Deutschland. Deshalb nutzte er die Gelegenheit eines Lehrauftrags an der Heinrich-Heine-Universität Düsseldorf, um sich gemeinsam mit seinen dortigen Geschichtsstudent*innen in einem Kurs auf die „Spurensuche [nach den] frühen Spiele-Entwickler*innen im deutschsprachigen Raum" zu begeben. Gemeinsam mit den Student*innen Constantin Bintz, Marlon Duncan Bonsch, Lars Brandes, Lisa Bresgott, Rika Bunse, Noah Dix, Victoria Hou, Daniel Kaspereit, Petros Kiorpes-Betchawas, Simon Körner, Rabea Kuschel, Christian Mischke,

Sebastian Müller, Tanja Pabst, Ann-Kristin Potthast, Deniz Sargin, Clarissa Schiffer, Jan Stockschläger und Ebru Yaylali sammelte er im Rahmen der Lehrveranstaltungen Titel für eine gemeinsame Datenbank in Form einer Google-Sheets-Tabelle. Zuerst wurden in Kleingruppen jene Plattformen durchsucht, die sich (zumindest über Umwege) nach Herkunftsländern filtern ließen: So wurde zum Beispiel auf den Plattformen Lemon Amiga und Interactive Fiction Database nach Spielen in deutscher Sprache gesucht.[5] In einem nächsten Schritt wurde dann auch auf anderen Plattformen nach Studios und einzelnen Entwickler*innen gesucht, die eindeutig dem DACH-Raum zugeordnet werden konnten. In der Folge wurden alle Listings des *Computronic*-Magazins aufgenommen, die über das Internet Archive verfügbar waren. Die Kleingruppen kontrollierten dabei jeweils die Eingaben anderer Kleingruppen. So entstand zuerst eine Alpha-Version und nach gemeinsamer Überarbeitung eine gesäuberte Beta-Version mit bereits ca. 600 Titeln, darunter aber nach wie vor einige Dubletten und Falschzuordnungen. Basierend auf dieser Liste haben die Student*innen dann historische Fallstudien einzelner Titel erstellt (z. B. Bonsch und Bunse 2023; Bresgott et al. 2022; Müller und Sargin 2023). Nach Abschluss des Semesters hat Ann-Kristin Potthast unter Mitarbeit von Marlon Duncan Bonsch, Lisa Bresgott, Rika Bunse, Clarissa Schiffer und Jan Stockschläger geholfen, die Datenbank gemeinsam mit Eugen Pfister und Lukas Daniel Klausner noch einmal gründlich zu kontrollieren. Zu jedem Titel wurden zusätzliche Quellen gesucht und angegeben. Doppelnennungen – zum Beispiel bedingt durch die Existenz mehrerer Namen für ein und dasselbe Spiel – wurden aufgelöst. Außerdem hat Eugen Pfister immer wieder freie Minuten genutzt, um nach weiteren Quellen, Studios und Entwickler*innen zu suchen.

## 4. Von der Beta- zur Gamma-Version

Mit Beginn des SNF-Sinergia-Projekts kamen dann die Projektmitglieder Adrian Demleitner, Addrich Mauch und Aurelia Brandenburg hinzu, welche die mittlerweile knapp 1200 Titel noch einmal zu kontrollieren halfen. Aufbauend auf der Beta-Version haben wir anschließend die Gamma-Version so entwickelt, wie sie inzwischen auch online frei abrufbar ist (Pfister et al. 2023). Dabei wurden wir mit denselben Problemen konfrontiert, wie sie häufig auftreten, wenn man sich darum bemüht, wenig aufgearbeitete und komplexe historische Daten unter einem bestimmten Aspekt zu strukturieren und zu säubern. So waren zum Beispiel die Genres der Spiele eine relevante Information und damit ein wichtiges Feld, auch um einen einzelnen Datensatz später einordnen zu können – aber gerade Genreeinteilungen sind meist fließend und schlecht zu standardisieren. Genauso waren wir mit einer großen Bandbreite an Qualität der recherchierbaren Informationen pro Titel konfrontiert: Für manche Spiele ließen sich verhältnismäßig leicht z. B. über MobyGames umfangreiche Listen der beteiligten Entwickler*innen recherchieren, die dann fast zu umfangreich für ein einzelnes Feld in der Datenbank waren, in anderen Fällen war auch mit genauerer Prüfung fürs Erste nur ein Studio oder ein Publisher, aber keine individuellen Entwickler*innen zu eruieren.

---

5 https://ifdb.org/search?sortby=old&searchfor=language%3Agerman (Suche nach Sprache, „earliest publications first") und https://www.lemonamiga.com/games/advanced_search.php (Advanced Search, suchen nach „Language: German").

Gerade Entwickler*innen (aber auch Publisher) boten auch über die bloße Informationsdichte hinaus einige Probleme in diesem Arbeitsschritt, weil beides immer wieder die Frage aufwarf, wann ein Spiel nun denn tatsächlich ein Spiel aus dem DACH-Raum war und wann nicht. In manchen Fällen lassen sich etwa einzelne Team-Mitglieder unter den Entwickler*innen (z. B. bei *Rayman* der Schweizer Yann Le Tensorer) ausmachen, die aus dem DACH-Raum stammen und so ein gutes Beispiel für eine internationale Vernetzung unter Entwickler*innen darstellen, während aber die Studios dieser Spiele ihren Sitz außerhalb Deutschlands, Österreichs oder der Schweiz hatten. Solche Nuancen verdeutlichen natürlich ganz automatisch, dass gerade in Hinblick auf digitale Spiele nationale Grenzen fließend sind und auch diese Art Sonderfälle sind (mit entsprechendem Vermerk) in der Gamma-Version aufgeführt – nur führen sie konkret vor Augen, wie schwer sich historische Daten Datenbankstrukturen unterwerfen lassen.

Dazu kam auch, dass wir uns durch unseren möglichst inklusiven Zugang der Erfassung (möglichst) aller Spiele zwar bewusst gängigen Kanonisierungsprozessen dessen, was heute meist als „Gamingkultur" gilt, widersetzt haben, dieser Zuschnitt aber auch dazu führt, dass auch die fertige Gamma-Version einige Titel enthält, über die wir über ihre bloße Existenz hinaus wenig Informationen erfassen konnten. In manchen Fällen lag das auch schon an der Publikationsform: Spiele wie *Time-Rally* (1987) wurden z. B. als Listings in Computerspielmagazinen – in diesem Fall der *Computronic* – veröffentlicht und entsprachen damit grundsätzlich unseren Kriterien, sind aber natürlich nicht ohne Weiteres im selben Maße verarbeitbar wie viele kommerziell vertriebene Spiele.

## 5. Ausblick

Wir sind uns dessen bewusst, dass unsere Datenbank – zum derzeitigen Zeitpunkt wie auch in absehbarer Zukunft – unvollkommen ist. So haben wir bisher nur die Listings aus einem Magazin aufgenommen. Wir müssen sogar davon ausgehen, dass selbst viele der kommerziell vertriebenen Spiele nach wie vor fehlen. Des Weiteren ist die Zuordnung zu Genres nach wie vor problematisch. Das liegt bis zu einem gewissen Grad in der Natur der Sache, weil es hierzu in der Forschung nach wie vor keinen Konsens gibt. Zugleich ist eine Zuordnung zu Genres gerade für unsere Forschung hoch relevant – denken wir nur an die eingangs gestellte Frage zur deutschsprachigen Sonderstellung des Genres Wirtschaftssimulation. Außerdem verschwinden in der derzeitigen

Tabellenstruktur die beteiligten Entwickler*innen, seien sie nun Programmierer*innen, Grafiker*innen, Autor*innen usw., in einer Kolonne der Tabelle.

Diese oft frustrierende Ungenauigkeit haben wir aber bewusst in Kauf genommen, eben weil wir uns entschieden haben, nicht auf eine Finanzierung zu warten, sondern über eine lange Zeit hinweg in unserer Freizeit voranzutreiben. Es war uns wichtig, möglichst bald erste Daten zur Verfügung zu stellen – nicht nur für unsere eigene Forschung, sondern für alle, die daran interessiert sind. Manchmal ist es auch in der Wissenschaft notwendig, „quick and dirty" vorzugehen, um die Forschung weiter anzustoßen. Es erscheint doch erstaunlich, dass es bis 2022 keine einschlägige, gut zugängliche Online-Datenbank für Game Studies unter dem Blickwinkel der *local history* gab. Und wir selbst haben schon jetzt enorm von der Arbeit profitiert – so ist das Sample von Schweizer Spielen für das SNF-Sinergia-Projekt von knapp über 50 auf 122 angestiegen, übrigens auch dank der Recherchen unserer Kolleg*innen aus dem Projekt. Auch zeigen sich schon in dem vorhandenen Sample an Spielen erste interessante Muster, die es weiter zu erforschen gälte. So fällt zum Beispiel auf, dass gerade in den 1980er- und 1990er-Jahren nur wenige Spiele in den größten deutschsprachigen Städten entwickelt wurden und kleine bzw. mittelgroße Städte hier überwiegen. Im selben Zeitraum finden sich in vielen Entwickler*innenteams öfter der gleiche Nachname, was auf Geschwister- oder Ehepaare schließen ließe. Außerdem kann man bereits sehr schön Konjunkturen einzelner Spielmechaniken nachzeichnen.

Der nächste Schritt wäre es, die Daten weiter zu säubern, sich der ungeliebten Genre-Frage zu stellen und allgemein saubere Metadatenstandards zu diskutieren. Dabei stehen wir aber erst am Anfang unserer Arbeit. Es ist unser Ziel, an der Datenbank beständig weiterzuarbeiten. Zugleich wünschen wir uns aber auch, dass möglichst viele von dieser Vorarbeit – denn mehr ist es bislang nicht – profitieren, weswegen wir schon die Gamma-Version trotz all ihrer Leerstellen und potenziellen Fehler im Rahmen einer CC-BY-SA-Lizenz online gestellt haben. Die Daten gehören nicht uns, sondern allen, die etwas damit anfangen können und wollen.

Parallel sind wir auch aktiv bemüht, die Ergebnisse in andere Datenbanken einzuarbeiten, zuallererst den bereits genannten Swiss Games Garden, aber auch – und hier für ein potenziell weitaus größeres Publikum – in das Wikimedia-Projekt Wikidata. Während der Integration unserer bisherigen Daten in Wikidata begegnen wir erneut ähnlichen Herausforderungen wie beim Aufbereiten der Liste selbst. Das WikiProject *Video Games*,[6] eine Arbeitsgruppe, die sich mit der Eintragung von digitalen Spielen in Wikidata befasst, hat ein eigenes Datenmodell entwickelt und im Zuge dessen Vorschläge für Genres und andere Attribute erstellt. Diese decken sich nicht immer mit unserer Liste, weshalb zusätzlicher Aufwand erforderlich ist und Konsense oder Kompromisse gefunden werden müssen. Wir nehmen während dieses Prozesses aktiv an der Arbeitsgruppe teil und haben uns auch bei Jean-Frédéric Berthelot, welcher seit Langem an dieser Arbeitsgruppe mitwirkt, um Unterstützung bemüht. Laut Berthelot hat Wikidata nicht den Anspruch, alle anderen Plattformen und Datenbanken zu ersetzen, sondern soll wie das Pilzmyzel im Wald[7] verschiedene Spieledatenbankprojekte miteinander verbinden. Daher ist die Verknüpfung zu den Einträgen eines Spiels auf anderen Seiten von hoher Bedeutung.

---

6 https://www.wikidata.org/wiki/Wikidata:WikiProject_Video_games

Das endgültige Ziel unseres Projekts ist es, verschiedenen Nutzer*innengruppen eine umfassende, möglichst vollständige und wissenschaftlich validierte Datenquelle zur Verfügung zu stellen. Nicht zuletzt wäre dies natürlich auch für journalistische Recherche dienlich, wenn sich Journalist*innen nicht auf unsichere Informationen von privaten Websites verlassen möchten. Es ist eine große Freude zu beobachten, wie viele Menschen bereits in den ersten Wochen nach Veröffentlichung auf unsere Datenbank reagierten. Teilweise kamen einzelne Ergänzungen herein, teilweise ganze Listen mit mehreren hundert Titeln, die es nun einzuarbeiten gilt. Vor allem aber freut uns, wie viele Kolleg*innen sich von unserer Arbeit inspiriert zeigen – und die Datenlage in Zukunft hoffentlich noch besser machen.

## Literatur

---

7 https://commonists.wordpress.com/2019/10/10/wikidata-the-underground-fungus-in-the-vast-forest-that-is-the-internet/